\documentclass[reprint, twocolumn, longbibliography, amsmath,amssymb, aps,]{revtex4}
\usepackage[breaklinks=true,colorlinks=true,linkcolor=blue,urlcolor=blue,citecolor=blue]{hyperref}
\usepackage{pdfpages}
\usepackage{textcomp}
\usepackage{amsmath} 
\usepackage{graphicx} 
\usepackage{float}
\usepackage{verbatim}
\usepackage{color} 
\usepackage{subfigure}  
\usepackage{hyperref}  
\usepackage[utf8]{inputenc}
\usepackage[english]{babel}
\usepackage{threeparttable}
\usepackage{amsmath}

\begin{document}

\title{A One-Dimensional Reduction Method for Calculating Thermal Expansion in Solids: Application to Orthorhombic Systems}

\author{Dmitry Vasilyev$^{*}$}
\affiliation{Baikov Institute of Metallurgy and Materials Science of the Russian Academy of Sciences}
\affiliation{D. Mendeleev University of Chemical Technology of Russia, Moscow, Russia}
\affiliation{HSE University, 101000 Moscow, Russia}
\email{vasilyev-d@yandex.ru}

\date{\today}

\begin{abstract}
Anisotropic thermal expansion plays a critical role in the performance and reliability of functional materials, yet its theoretical description remains limited. Here, a computational framework that reduces the calculation of thermal expansion in solids to an effective one-dimensional problem is presented and applied to orthorhombic Mo$_2$C. The approach explicitly incorporates electronic, vibrational, and magnetic contributions to the free energy. Using this method, a comprehensive set of thermodynamic and mechanical properties is determined, including elastic constants, bulk, shear, and Young’s moduli, Debye temperature, Poisson’s ratio, sound velocities, elastic anisotropy, and heat capacity. The predicted properties are in good agreement with available experimental data. The results reveal pronounced thermal expansion anisotropy in orthorhombic Mo$_2$C and provide quantitative insight into the thermoelastic behavior of orthorhombic-phase materials, with implications for the design and optimization of Mo$_2$C-based catalytic systems.
\end{abstract}

\maketitle


\section{Introduction}

A significant challenge arises from the fact that the volume of the orthorhombic lattice is dependent on three parameters—namely, $a$, $b$, and $c$. Consequently, the optimization of these parameters and the calculation of the thermodynamic free energy function, $F(a,b,c,T)$, must be performed considering all three variables. This poses a mathematical challenge in accurately capturing the behavior of the material. Because equations composed as functions $F(a,b,c,T)$ to describe the thermodynamics of a solid in the mathematical sense are transcendental equations. Currently, there is no general solution to transcendental equations. Only particular solutions can be found. To address this, a modified Search of Thermal Expansion Path (STEP) method~\cite{Vasilyev2021PhysicaB} ~\cite{Vasilyev2023MTComm} ~\cite{Vasilyev2023PSSB} ~\cite{Vasilyev2024PCCP} ~\cite{VasilyevGorev2025PhysicaB} has been employed, which simplifies the optimization process by reducing the dimensionality of the problem. This method explores the thermal expansion along various crystallographic directions, transforming the problem into a one-dimensional case dependent solely on the volume, V. This simplification allows for the application of classical thermodynamic relations, including the Murnaghan equation of state, the quasi-harmonic Debye–Grüneisen approximation (QHA), electronic, vibrational and magnetic entropies, which are now functions of only one variable, V. First-principles density functional theory (DFT) calculations are used to account for both electronic and vibrational contributions to the system.

In this study, the free energies of orthorhombic $\alpha$-Mo$_2$C were computed along multiple thermal-expansion pathways, with the equilibrium thermal-expansion path being selected based on the minimum free energy trajectory. Results indicate significant anisotropy in the thermal expansion of $\alpha$-Mo$_2$C. The contributions of electronic, vibrational entropies to the material’s phase stability were thoroughly examined. Additionally, key thermodynamic properties such as isobaric heat capacity, thermal-expansion coefficients, elastic constants, bulk modulus, sound velocities, Debye temperature, and lattice thermal conductivity were determined. The obtained results are consistent with available experimental data, providing a deeper understanding of the thermal-expansion behavior of orthorhombic phases and offering valuable insights for the future development of advanced materials.

\begin{figure}[h]
\centering
  \includegraphics[width=8.5cm]{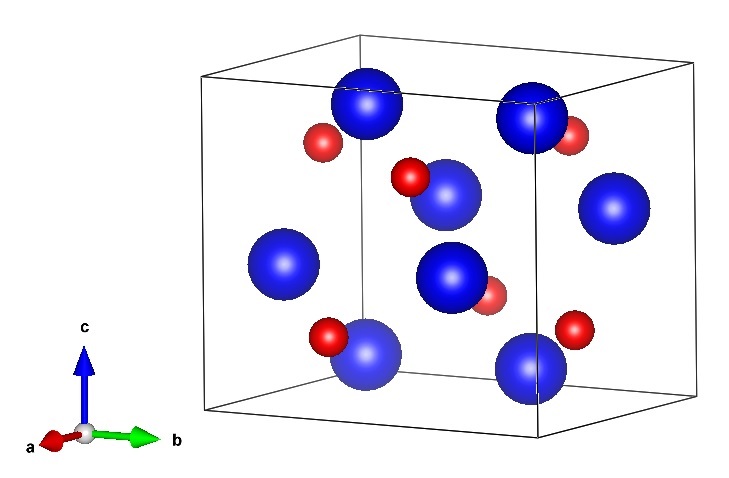}\\
  \caption 
  {Schematic representation of the unit cell of the orthorhombic Mo$_2$C phase, where Mo atoms at ($8d$) sites are shown in blue, and C atoms at ($4c$) sites are shown in red}
  \label{fig:cell}
  \end{figure}

Figure~\ref{fig:cell} depicts the crystal lattice of $\alpha$-Mo$_2$C, which crystallizes in an orthorhombic structure ($Pbcn$ space group, No. 60). In this schematic, blue spheres represent Mo atoms occupying ($8d$) Wyckoff positions, while red spheres represent C atoms located at the ($4c$) sites.

\begin{figure*}[t]
\centering
  \includegraphics[width=11cm]{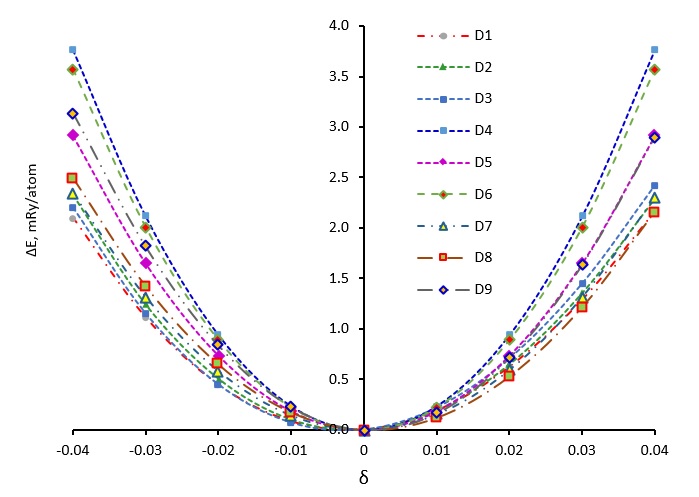}\\
  \caption 
  {Graphs of the change in the total energy $\Delta E = E_{tot}(V_0,\delta_i) - E_{tot}(V_0,0)$, obtained for different distortion matrices D1 - D9 as functions of the applied deformations ($\delta$) of the crystal lattice of $\alpha$-Mo$_2$C, calculated for the ground state. Dashed lines are polynomial approximations.}
\label{fig:grafs}  
\end{figure*}

\begin{table*}[t]
 \caption{The lattice constants (in Å), equilibrium volume V (in Å$^{3}$), formation enthalpies $\Delta H$ (eV/atom) of orthorhombic $\alpha$-Mo$_2$C phase calculated at T = 0K presented with experimental and theoretical data for comparison.}
\begin{threeparttable}[b] 
 \label{table:f_Lattice_const}
\begin{tabular}{||c c c c c c c c c||} 
 \hline
 Method & Ref. & $a$ & $b$ & $c$ & $c/a$ & $c/b$ & $V$ & $\Delta H$\\ [1ex]
 \hline\hline
Calc. & This work & 4.750 & 6.069 & 5.235 & 1.102 & 0.863 & 150.91 & -0.19 \\ [1ex]
 \hline
 Calc. & This work$^{a}$ & 4.845 & 6.107 & 5.185 & 1.070 & 0.849 & 153.42 & - \\ [1ex]
 \hline
 Calc. & ~\cite{Politi2013PCCP} & 4.741 & 6.060 & 5.227 & 1.103 & 0.863 & 150.17 & -0.3 \\ [1ex]
 \hline
 Calc. & ~\cite{Vojvodic2012CatalLett} & 4.825 & 6.162 & 5.304 & 1.099 & 0.861 & 157.70 & - \\ [1ex]
 \hline
 Calc. & ~\cite{Medford2012JournalCatalysis} & 4.839 & 6.173 & 5.322 & 1.100 & 0.862 & 158.97 & -0.21 \\ [1ex]
 \hline
 Calc. & ~\cite{Haines2001JPCM} & 4.735 & 6.125 & 5.260 & 1.111 & 0.859 & 152.55 & - \\ [1ex]
 \hline
 Exp. & ~\cite{Otani1995JCG} & 4.729 & 6.028 & 5.197 & 1.099 & 0.862 & 149.05 & -0.49$^{e}$ \\ [1ex]
 \hline
 Calc. & ~\cite{Liu2009ScriptaMater} & 4.738 & 6.038 & 5.210 & 1.100 & 0.863 & 148.63 & - \\ [1ex]
 \hline
 Exp. & ~\cite{Epicier1988ActaMet}$^{d}$ & 4.735 & 6.025 & 5.210 & 1.100 & 0.865 & 147.82 & - \\ [1ex]
 \hline
 Exp. & ~\cite{Naher2022ResultsPhys} & 4.725 & 6.022 & 5.195 & 1.099 & 0.863 & 148.83 & - \\ [1ex]
 \hline
 Calc. & ~\cite{GuardiaValenzuela2018Carbon}$^{b}$ & 4.739 & 6.022 & 5.215 & 1.100 & 0.866 & 147.98 & - \\ [1ex]
 \hline
 Exp. & ~\cite{Ge2021JAlloysCompd}$^{c}$ & 4.733 & 6.007 & 5.205 & 1.1000 & 0.866 & 147.97 & - \\ [1ex]
 \hline
 Calc. & ~\cite{Naher2022ResultsPhys} & 4.705 & 6.008 & 5.172 & 1.099 & 0.861 & 146.20 & - \\ [1ex]
 \hline
 Calc. & ~\cite{Karaca2019JAlloysCompd} & 4.75 & 6.09 & 5.26 & 1.107 & 0.864 & 152.16 & - \\ [1ex]
 \hline
 Exp. & ~\cite{Karaca2019JAlloysCompd}$^{f}$ & 4.75 & 6.03 & 5.23 & 1.101 & 0.867 & 149.80 & - \\ [1ex]
 \hline
 Exp. & ~\cite{Reddy2010JAlloysCompd}$^{g}$ & 4.748 & 6.020 & 5.213 & 1.098 & 0.866 & 149.00 & - \\ [1ex]
 \hline
 Exp. & ~\cite{Parthe1963ActaCryst}$^{h}$ & 4.724 & 6.004 & 5.199 & 1.101 & 0.866 & 147.46 & - \\ [1ex]
 \hline
\end{tabular}
    \begin{tablenotes}
      \item[a] Calculated in this work at T = 1000 K.
      \item[b] Calculated from hexagonal molybdenum sublattice parameters.
      \item[c] Crystal details obtained from Rietveld refinement.
      \item[d] Neutron diffraction data. The samples were processed at a pressure from 15 to 49 GPa and a temperature of 1000 $^{o}$C
      \item[e] The free energy of formation of $\alpha$-Mo$_2$C was determined in the temperature range from 1270 to 1573 K using the carbon activity in the two-phase region Mo + Mo$_2$C ~\cite{Seigle1979MetallTransA}.
      \item[f] experimentaly obtaibed at T = 800 $^{o}$C.
      \item[g] Crystalline $\alpha$-Mo$_2$C was obtained on heat treatment of a C-rich precursor at 1000 $^{o}$C after 1 h of isothermal holding under hydrogen atmosphere.
      \item[h] The data were obtained from neutron diffraction powder pattern of Mo$_2$C.
    \end{tablenotes}
  \end{threeparttable}
\end{table*}

\begin{table}[t]
 \caption{Calculated atomic coordinates of the $\alpha$-Mo$_2$C compound in the crystal lattice for the ground state.}
 \label{table:struct_param}
\begin{tabular}{||c c c c c||} 
 \hline
 Atoms & Wyckoff & x & y & z \\ [0.5ex] 
 \hline\hline
 Mo & \textit{8d} & 0.2470 & 0.1230 & 0.0803  \\ [1ex] 
 \hline
 Mo & \textit{8d} & 0.7530 & 0.8770 & 0.9197  \\ [1ex]
 \hline
 Mo & \textit{8d} & 0.7470 & 0.6230 & 0.4197 \\ [1ex]
 \hline
 Mo & \textit{8d} & 0.2530 & 0.3770 & 0.5803 \\ [1ex]
 \hline
 Mo & \textit{8d} & 0.2530 & 0.6230 & 0.0803 \\ [1ex]
 \hline
 Mo & \textit{8d} & 0.7470 & 0.3770 & 0.9197 \\ [1ex]
 \hline
 Mo & \textit{8d} & 0.7530 & 0.1230 & 0.4197 \\ [1ex]
 \hline
 Mo & \textit{8d} & 0.2470 & 0.8770 & 0.5803 \\ [1ex]
 \hline\hline
 C & \textit{4c} & 0 & 0.3768 & 0.25 \\ [1ex]
 \hline
 C & \textit{4c} & 0 & 0.6232 & 0.75 \\ [1ex]
 \hline
 C & \textit{4c} & 0.5 & 0.8768 & 0.25 \\ [1ex]
 \hline
 C & \textit{4c} & 0.5 & 0.1232 & 0.75 \\ [1ex]
 \hline
\end{tabular}
\end{table}
\begin{table*}[t]
 \caption{Distortion matrices D$_i$ and the system of equations describing the change in the total energy $\Delta$E($\delta$) of the orthorhombic compound after lattice deformation by $\delta$. Explanations are given in the text.}
 \label{table:params}
\begin{tabular}{|c c|}
\hline
 D$_i$ & $\Delta$E($\delta$) \\ [0.5ex] 
 \hline\hline
  $D_1 = \begin{pmatrix}
1 + \delta \ & 0 & 0\\
0 & 1 & 0\\
0 & 0 & 1
\end{pmatrix}$ & $\Delta E = V_0 \cdot (\tau 1 \cdot \delta + \frac{C11}{2} \cdot \delta^2)$ \\ 
 \hline
   $D_2 = \begin{pmatrix}
1 & 0 & 0\\
0 & 1 + \delta\ & 0\\
0 & 0 & 1 
\end{pmatrix}$ & $\Delta E = V_0 \cdot (\tau 2 \cdot \delta + \frac{C22}{2} \cdot \delta^2)$\\
 \hline
   $D_3 = \begin{pmatrix}
1 & 0 & 0\\
0 & 1 & 0\\
0 & 0 & 1 + \delta\
\end{pmatrix}$ & $\Delta E = V_0 \cdot (\tau 3 \cdot \delta + \frac{C33}{2} \cdot \delta^2)$\\
 \hline
   $D_4 = \begin{pmatrix}
\frac{1}{(1-\delta^2)^\frac{1}{3}} & 0 & 0\\
0 & \frac{1}{(1-\delta^2)^\frac{1}{3}} & \frac{\delta}{(1-\delta^2)^\frac{1}{3}}\\
0 & \frac{\delta}{(1-\delta^2)^\frac{1}{3}} & \frac{1}{(1-\delta^2)^\frac{1}{3}}
\end{pmatrix}$ & $\Delta E = V_0 \cdot (2 \cdot \tau 4 \cdot \delta + 2 \cdot C44 \cdot \delta^2)$\\ [1ex] 
 \hline
   $D_5 = \begin{pmatrix}
\frac{1}{(1-\delta^2)^\frac{1}{3}} & 0 & \frac{\delta}{(1-\delta^2)^\frac{1}{3}}\\
0 & \frac{1}{(1-\delta^2)^\frac{1}{3}} & 0\\
\frac{\delta}{(1-\delta^2)^\frac{1}{3}} & 0 & \frac{1}{(1-\delta^2)^\frac{1}{3}}
\end{pmatrix}$ & $\Delta E = V_0 \cdot (2 \cdot \tau 5 \cdot \delta + 2 \cdot C55 \cdot \delta^2)$\\ [1ex] 
 \hline
    $D_6 = \begin{pmatrix}
\frac{1}{(1-\delta^2)^\frac{1}{3}} & \frac{\delta}{(1-\delta^2)^\frac{1}{3}} & 0\\
\frac{\delta}{(1-\delta^2)^\frac{1}{3}} & \frac{1}{(1-\delta^2)^\frac{1}{3}} & 0\\
0 & 0 & \frac{1}{(1-\delta^2)^\frac{1}{3}}
\end{pmatrix}$ & $\Delta E = V_0 \cdot (2 \cdot \tau 6 \cdot \delta + 2 \cdot C66 \cdot \delta^2)$\\ [1ex] 
 \hline
   $D_7 = \begin{pmatrix}
\frac{1+\delta}{(1-\delta^2)^\frac{1}{3}} & 0 & 0\\
0 & \frac{1-\delta}{(1-\delta^2)^\frac{1}{3}} & 0\\
0 & 0 & \frac{1}{(1-\delta^2)^\frac{1}{3}}
\end{pmatrix}$ & $\Delta E = V_0 \cdot \big((\tau 1 -\tau 2) \delta + \frac{1}{2}(C11+C22-2 \cdot C12) \delta^2\big)$ \\
 \hline 
   $D_8 = \begin{pmatrix}
\frac{1+\delta}{(1-\delta^2)^\frac{1}{3}} & 0 & 0\\
0 & \frac{1}{(1-\delta^2)^\frac{1}{3}} & 0\\
0 & 0 & \frac{1-\delta}{(1-\delta^2)^\frac{1}{3}}
\end{pmatrix}$ & $\Delta E = V_0 \cdot \big((\tau 1 -\tau 3) \delta + \frac{1}{2}(C11+C33-2 \cdot C13) \delta^2\big)$\\
 \hline 
   $D_9 = \begin{pmatrix}
\frac{1}{(1-\delta^2)^\frac{1}{3}} & 0 & 0\\
0 & \frac{1+\delta}{(1-\delta^2)^\frac{1}{3}} & 0\\
0 & 0 & \frac{1-\delta}{(1-\delta^2)^\frac{1}{3}}
\end{pmatrix}$ & $\Delta E = V_0 \cdot \big((\tau 2 -\tau 3) \delta + \frac{1}{2}(C22+C33-2 \cdot C23) \delta^2\big)$\\
 \hline 
\end{tabular}
\end{table*}

\section{Calculations}

First-principles calculations were performed using the WIEN2k software package~\cite{Blaha2020JCP}, employing the full-potential linearized augmented plane-wave (FP-LAPW) method~\cite{Perdew1996PRB}. Exchange–correlation effects were treated within the generalized gradient approximation (GGA) using the Perdew–Burke–Ernzerhof (PBE) functional. The Brillouin zone was sampled using a Monkhorst–Pack k-point mesh of 11 × 8 × 10 in the first irreducible Brillouin zone~\cite{Monkhorst1976PRB}. Structural optimizations were carried out until convergence criteria of less than 0.5 mRy/Bohr for atomic forces, below 0.01 GPa for residual pressure, and 10$^{-6}$ Ry/atom for total energy were satisfied.

\section{Ground state properties of M$o$$_2$C}

The total energy of the orthorhombic Mo$_2$C compound was obtained through first-principles calculations. To determine the lattice parameters and atomic positions of the Mo$_2$C compound, geometry optimization procedures and full structural relaxation were employed.

The optimized lattice parameters, the ratios of ($c/a$) and ($c/b$), and the resulting volume V of the Mo$_2$C compound are presented in Table ~\ref{table:f_Lattice_const}, alongside other experimental and calculated values from previous studies. The enthalpy of formation of Mo$_2$C, calculated according to Equation (1), relative to the body-centered cubic (bcc) structure of molybdenum and carbon in the graphite structure, is also presented in Table~\ref{table:f_Lattice_const}:
\begin{equation}
    \Delta H^{Mo2C} = E_{tot}^{Mo2C} - (x \cdot E_{graphite}^C +(1 - x) \cdot E_{bcc}^{Mo})
\end{equation}

where E$^{Mo2C}$$_{total}$ is the total energy of $\alpha$-Mo$_2$C, E$^{Mo}$$_{bcc}$ and E$^{C}$$_{graphite}$ are the energies of Mo atoms in the bcc lattice and C atoms in the graphite structure, respectively, and x represents the concentration of carbon atoms (x = 0.333). The calculated value of $\Delta$H$^{Mo2C}$, presented in this study, is negative, indicating that the compound is thermodynamically stable in its ground state. Another value for $\Delta$H was obtained in the work of Seigle et al.~\cite{Seigle1979MetallTransA}, where the authors used the measured carbon activity in the two-phase region of Mo + Mo$_2$C within a temperature range of 1270–1573 K. This and other values of $\Delta$H, calculated by different authors, are listed in Table~\ref{table:f_Lattice_const} for comparison.

The optimization results revealed the following crystal lattice parameters: a = 4.750 Å, b = 6.069 Å, and c = 5.235 Å. The coordinates of the relaxed atoms are provided in Table~\ref{table:struct_param}, and the distribution of atoms across the sublattices—eight molybdenum atoms in Wyckoff positions 8d and four carbon atoms in positions 4c—are shown in Figure~\ref{fig:cell}.

\section{Elastic properties of a single crystal of the orthorhombic phase $\alpha$-Mo$_2$C}

In the regime of small elastic strains, where the generalized Hooke's law can be applied, the relationship between the strain $\sigma_{i}$ in the crystal lattice of a solid under an external load and the stresses $\epsilon_j$ can be expressed as follows:

\begin{equation}
    \sigma_{i} = \sum_{j=1}^{6} C_{ij}\epsilon_{j} 
\end{equation}
where $C_{ij}$ are the coefficients of the elastic deformation tensor. In turn, the stress can be expressed as a function of strain. This leads to the following equation:
\begin{equation}
    \epsilon_{i} = \sum_{j=1}^{6} S_{ij}\sigma_{j}
\end{equation}
where $S_{ij}$ are the elements of the compliance matrix. In general, the internal energy of a deformed crystal can be expanded as a Taylor series with respect to the stress tensor:
\begin{equation}
    E(V,(\epsilon_{i})) = E(V_{0},0) + \frac{V_{0}}{2} \sum_{i,j=1}^{6} C_{ij}\epsilon_{i}\epsilon_{j} + ...,
\end{equation}
where E(V$_0$,0) and V$_0$ represent the energy and volume of the equilibrium structure in its initial state, respectively.

To calculate the elastic constants $C_{ij}$, a small external stress is applied to the equilibrium lattice, and the corresponding change in the total energy is determined. The elastic deformation energy is given by the following expression~\cite{Nye1964Crystals}:
\begin{equation}
    \Delta E = \frac{V_{0}}{2} \sum_{i,j=1}^{6} C_{ij}\epsilon_{i}\epsilon_{j},
\end{equation}
where $\Delta E = E_{tot}(V_0,\delta_i) - E_{tot}(V_0,0)$ is the total energy difference between the deformed and original lattices, V$_0$ is the volume of the equilibrium unit cell, and $C_{ij}$ are the elastic constants; $\epsilon_i$ and $\epsilon_j$ represent strains, and $\delta$ is the deformation added to the equilibrium lattice.

The deformation tensor $C_{ij}$ for the Mo$_2$C compound was calculated based on the changes in the total energy $\Delta$E of the crystal lattice resulting from the sequential application of distortion matrices \textbf{D$_i$} listed in Table~\ref{table:params}, according to the matrix equation:
\begin{equation}
\label{eqe:mtrx_eq}
    \textbf{R} \cdot\ \textbf{D$_i$} = \textbf{R'}
\end{equation}
\begin{equation}
\begin{pmatrix}
\frac{1}{\sqrt 2} & \frac{1}{\sqrt 2} & 0\\
\frac{- 1}{\sqrt 2} & \frac{1}{\sqrt 2} & 0\\
0 & 0 & 1
\end{pmatrix}
\end{equation}
where \textbf{R} is the matrix defined by the Bravais lattice vectors of the orthorhombic crystal, representing the original lattice of the phase under study; \textbf{R'} is the distorted matrix.

As a result, the original Mo$_2$C lattice, represented by the matrix \textbf{R}, is deformed. By solving the matrix equation~\ref{eqe:mtrx_eq}, the deformed Mo$_2$C lattice is described by the matrix \textbf{R'}. The energy differences $\Delta E (\delta) = E_{tot}(V_0,\delta_{i}) - E_{tot}(V_0,0)$ for each distortion are calculated and presented as functions $\Delta E(\delta)$ in Figure ~\ref{fig:grafs}. By solving the system of equations shown in Table~\ref{table:params}, the coefficients of the deformation tensor C$_{ij}$ can be determined. The nine elastic constants C$_{ij}$ calculated are given in Table~\ref{table:elstconst}.

To assess whether the investigated compound exhibits mechanical stability, the generalized criterion for the mechanical stability of orthorhombic crystals at zero pressure, as described in ~\cite{Wu2007PhysRevB}, can be applied. This criterion imposes the following conditions on the elastic constants at zero pressure:
\begin{equation}
\begin{aligned}
& (C_{11} + C_{22} - 2C_{12}) > 0, \\
& (C_{22} + C_{33} - 2C_{23}) > 0, \\
& (C_{11} + C_{33} - 2C_{13}) > 0, \\
& C_{ii} > 0 \quad (i = 1,2,\ldots,6), \\
& (C_{11} + C_{22} + C_{33} + 2C_{12} + 2C_{13} + 2C_{23}) > 0.
\end{aligned}
\label{eq:stability}
\end{equation}

The elastic constants C$_{ij}$ presented in Table~\ref{table:elstconst} show that this criterion is satisfied, indicating that Mo$_2$C is a mechanically stable compound at zero pressure and T = 0 K. The high values of the calculated elastic constants C$_{11}$, C$_{22}$ and C$_{33}$ suggest that resistance to compression along the $a$, $b$ and $c$ axes will be significant. The values of the elastic constants C$_{44}$, C$_{55}$ and C$_{66}$ describe the resistance to shear in the \{100\} plane along the $<110>$ direction, and in the \{010\} plane along the $<001>$ direction, respectively.

\begin{table*}[t]
 \caption{Elastic constants C$_{ij}$ (in GPa) of $\alpha$-Mo$_2$C, calculated for the ground state and obtained in this work, compared with results from different studies.}
 \label{table:elstconst}
\begin{tabular}{||c c c c c c c c c c||} 
 \hline
 Ref. & C$_{11}$ & C$_{22}$ & C$_{12}$ & C$_{13}$ & C$_{23}$ & C$_{33}$ & C$_{44}$ & C$_{55}$& C$_{66}$ \\ [0.5ex] 
 \hline\hline
This work & 460.66 & 497.93 & 224.14 & 223.92 & 160.34 & 500.07 & 203.65 & 160.10 & 193.29 \\ [1ex] 
 \hline
~\cite{Karaca2019JAlloysCompd} & 460.04 & 495.17 & 228.67 & 238.67 & 185.77 & 492.45 & 140.61 & 161.81 & 183.25 \\ [1ex] 
 \hline 
~\cite{Persson2015MaterialsProject} & 488 & 489 & 169 & 226 & 216 & 453 & 179 & 167 & 139 \\ [1ex] 
 \hline
~\cite{Naher2022ResultsPhys} & 485.84 & 515.32 & 236.61 & 235.42 & 182.95 & 518.37 & 155.27 & 178.66 & 197.62 \\ [1ex] 
 \hline 
\end{tabular}
\end{table*}

\section{Elastic properties of the polycrystalline orthorhombic phase}

The polycrystalline elastic properties of Mo$_2$C, including the bulk modulus ($B$), shear modulus ($G$), Young’s modulus ($E$), and Poisson’s ratio ($\nu$), were evaluated using the Voigt–Reuss–Hill (VRH) averaging scheme~\cite{Hill1952ProcPhysSoc}:

\begin{equation}
    B = \frac{1}{2}\Big(B_{V}+B_{R}\Big), G = \frac{1}{2}\Big(G_{V}+G_{R}\Big) 
\end{equation}
\begin{equation}
    E = \frac{9G\cdot B}{3B+G}, \nu = \frac{3B-2G}{2(3B+G)} 
\end{equation}

Here, B$_V$ and B$_R$ denote the Voigt and Reuss bounds of the bulk modulus, respectively, which represent the upper and lower limits of elastic response. The elastic moduli of Mo$_2$C were calculated following the formalism described in Ref.~\cite{Voigt1928Kristallphysik},~\cite{Reuss1929ZAMM}:

\begin{equation}
    B_{V} = \frac{1}{9}\Big(C_{11}+C_{22}+C_{33}+C_{12}+C_{13}+C_{23}\Big) 
\end{equation}
\begin{equation}
    B_{R} = \frac{1}{(S_{11}+S_{22}+S_{33})+2(S_{12}+S_{13}+S_{23})} 
\end{equation}

\begin{equation}
\begin{aligned}
G_V
&= \frac{1}{15} \left( C_{11} + C_{22} + C_{33} - C_{12} - C_{13} - C_{23} \right) \\
&\quad + \frac{1}{5} \left( C_{44} + C_{55} + C_{66} \right)
\end{aligned}
\end{equation}

%
\begin{equation}
\begin{aligned}
G_R
&= 15/\big(4(S_{11}+S_{22}+S_{33}) \\
&\quad - 4(S_{12}+S_{13}+S_{23}) \\
&\quad + 3(S_{44}+S_{55}+S_{66})\big)
\end{aligned}
\end{equation}

In these expressions, C$_{ij}$ are the single-crystal elastic stiffness constants, and S$_{ij}$ are the corresponding compliance coefficients obtained from the inverse of the stiffness tensor.

Using the calculated single-crystal elastic constants, the effective elastic properties of the polycrystalline aggregate were derived according to Eqs. (9–14). The resulting values of $B$, $G$, $E$, and $\nu$, calculated at T = 0 K, are summarized in Table~\ref{table:modulus}. The compliance constants S$_{ij}$ used in Eq. (14) are listed in Table~\ref{table:compliance}.

\begin{table}[t]
\centering
\caption{Compliance constants $S_{ij}$ (TPa$^{-1}$) for Mo$_2$C single-crystals at T = 0 K.}
\label{table:compliance}
\begin{tabular}{|c c c c c c c c c|} 
 \hline
 $S_{11}$ & $S_{22}$ & $S_{12}$ & $S_{13}$ &$S_{23}$ & $S_{33}$ & $S_{44}$ & $S_{55}$ & $S_{66}$ \\ [0.5ex] 
 \hline
 3.24 & 2.62 & -1.11 & -1.10 & -0.34 & 2.60 & 4.91 & 6.25 & 5.17 \\ [1ex] 
 \hline
\end{tabular}
\end{table}

The ductile or brittle character of Mo$_2$C was assessed using the criteria proposed by Pugh~\cite{Pugh1954PhilMag} and Lewandowski et al.~\cite{Lewandowski2005PhilMagLett}. According to Pugh’s criterion, materials with a $B/G$ ratio exceeding 1.75 are considered ductile. Similarly, a Poisson’s ratio greater than 0.26 has been suggested as indicative of ductile behavior. As shown in Table ~\ref{table:modulus}, the calculated $B/G$ and $\nu$ values for Mo$_2$C satisfy both criteria, indicating that Mo$_2$C exhibits ductile mechanical behavior under applied stress.

\section{Thermal properties of the compound}
The Debye temperature, $\theta_D$ , was determined from the bulk modulus $B$ and shear modulus $G$ of the polycrystalline aggregate via the average sound velocity~\cite{Anderson1963JPhysChemSolids}, using the transverse (shear) ($v_s$) and longitudinal ($v_l$ ) elastic wave velocities calculated according to:

\begin{equation}
    v_{s} = \Big(\frac{G}{\rho}\Big)^{1/2}, v_{l} = \Big(\frac{B + 4G/3}{\rho}\Big)^{1/2}
\end{equation}
where $B$ and $G$ are the bulk and shear moduli, respectively, and $\rho$ is the density of the compound. The average sound velocity $V_m$  in the polycrystal is given by:

\begin{equation}
    Vm = \Big[\frac{1}{3}(\frac{2}{vt^3} + \frac{1}{vl^3})\Big]^{-1/3}
\end{equation}
The Debye temperature $\theta_{D}$  was then evaluated using the expression adopted from Ref.~\cite{Anderson1963JPhysChemSolids}:

\begin{equation}
    \theta_D = \frac{h}{k_B}\Big[\frac{3n}{4\pi}\Big(\frac{N_{A} \rho}{M}\Big)\Big]^{1/3} \cdot V_M
\end{equation}
The propagation velocities of elastic waves in a single crystal along different crystallographic directions were derived from the calculated elastic constants by solving the Christoffel equation~\cite{Giurgiutiu2022AerospaceComposites}:

\begin{equation}
    |C_{ijkl}n_{j}n_{l} - \rho V^2 \delta_{ik}| = 0
\end{equation}
where C$_{ijkl}$ are the elastic stiffness tensor components of the single crystal, \textit{n} denotes the propagation direction, $\rho$ is the density, \textit{V} is the elastic wave velocity, and $\delta_{ik}$  is the Kronecker delta. This equation yields three solutions corresponding to one longitudinal wave velocity (V$_L$ ) and two transverse wave velocities (V$_S$$_1$ and V$_S$$_2$).

The sound velocities propagating along specific crystallographic directions in the orthorhombic crystal were calculated following Ref.~\cite{Brugger1965JAP}. For wave propagation along the [001] direction, the elastic wave velocities were obtained as:

\begin{equation}
    V_{L} = \Big(\frac{C_{33}}{\rho}\Big)^{1/2}, V_{S1} = V_{S2} = \Big(\frac{C_{44}}{\rho}\Big)^{1/2}
\end{equation}
Similarly, the elastic wave velocities along the [100] direction were calculated as:
\begin{equation}
    V_{L} = \Big(\frac{C_{11}}{\rho}\Big)^{1/2}, V_{S1} = \Big(\frac{C_{66}}{\rho}\Big)^{1/2}, V_{S2} = \Big(\frac{C_{44}}{\rho}\Big)^{1/2}
\end{equation}
where V$_L$ and V$_S$  denote the longitudinal and transverse (shear) elastic wave velocities, respectively.

The calculated transverse (V$_s$), longitudinal (v$_l$), and average (V$_M$ ) sound velocities, the Debye temperature $\theta_{D}$ obtained from Eqs. (15–17), and the anisotropic elastic wave velocities derived from Eqs. (19) and (20) for Mo$_2$C are summarized in Table~\ref{table:wave_vel}.

\begin{table}[t]
 \caption{Calculated average (V$_m$), transverse (v$_s$), and longitudinal (v$_l$) elastic wave velocities ($\mathrm{m\ s^{-1}}$); Debye temperature $\theta_{D}$ ($\mathrm{K}$); and elastic wave velocities along the [001] and [100] crystallographic directions ($\mathrm{m\ s^{-1}}$) for Mo$_2$C and their comparison with available theoretical data.}
 \label{table:wave_vel}
\begin{tabular}{||c c c c c c c c c c||} 
 \hline
 Compound & v$_s$ & v$_l$ & V$_m$ & $\theta_{D}$ & [001] & -- & [100] & -- & -- \\ [0.5ex] 
 \hline
 -- & -- & -- & -- & -- & V$_l$ & V$_s$ & V$_l$ & V$_{S1}$ & V$_{S2}$  \\ [0.5ex] 
 \hline\hline
This work & 4284 & 7588 & 4765 & 610 & 7465 & 4764 & 7165 & 4641 & 4764 \\ [1ex] 
 \hline
Calc. ~\cite{Karaca2019JAlloysCompd} & 4079 & 7522 & 4552 & 581 & - & - & - & - & - \\ [1ex] 
 \hline
Ref. ~\cite{Karaca2019JAlloysCompd} & 4146 & 7397 & 46125 & 593 & - & - & - & - & - \\ [1ex] 
 \hline
Exp. ~\cite{Cankurtaran2004JMaterSci} & 3905 & 6906 & 4343 & 559 & - & - & - & - & - \\ [1ex] 
 \hline  
\end{tabular}
\end{table}

The results reveal pronounced anisotropy in the elastic wave propagation velocities, as shown in Table~\ref{table:wave_vel}. The highest wave velocity is observed for longitudinal waves propagating along the [001] direction. The predicted directional dependence of elastic wave velocities may provide valuable guidance for future experimental investigations of the elastic and vibrational properties of Mo$_2$C.

\section{Difficulties in modeling the thermal expansion of solids}
To model thermodynamic functions and thermal expansion using first-principles calculations, the quasi-harmonic approximation (QHA) is commonly employed in materials science. Within this framework, the Helmholtz free energy is expressed as follows~\cite{Dove1993LatticeDynamics}:
\begin{equation}
\begin{split}
F(V,T) =\; &E_{\text{tot}}(V) + F_{\text{el}}(V,T) + F_{\text{vib}}(V,T) \\
           &+ F_{\text{mag}}(V,T) - T S_{\text{conf}}
\end{split}
\end{equation}
where E$_{tot}$(V) represents the total energy derived from DFT calculations at T = 0 K. Additional contributions to the free energy include the electronic, vibrational, magnetic components, and configurational entropy.

The thermal electron energy, F$_{el}$(V,T), is calculated in accordance with the formula provided in~\cite{LandauLifshitz1980StatPhys}:
\begin{equation}
    F_{el}(V,T) = E_{el}(V,T) - TS_{el}(V,T) 
\end{equation}
where the electronic entropy, S$_{el}$(V,T), is given by:

\begin{equation}
\begin{split}
S_{\mathrm{el}}(V,T)
= -k_{\mathrm{B}} \int_{-\infty}^{\infty}
n(\varepsilon,V)
\Bigl[
f(\varepsilon,T)\ln f(\varepsilon,T) \\
\qquad + \bigl(1 - f(\varepsilon,T)\bigr)
\ln\bigl(1 - f(\varepsilon,T)\bigr)
\Bigr]
\, d\varepsilon
\end{split}
\end{equation}

The electronic energy E$_{el}$(V,T) takes the form:

\begin{equation}
\begin{split}
E_{\mathrm{el}}(V,T)
= \int_{-\infty}^{\infty}
n(\varepsilon,V)\, f(\varepsilon,T)\, \varepsilon \, d\varepsilon \\
\qquad - \int_{-\infty}^{\varepsilon_F}
n(\varepsilon,V)\, \varepsilon \, d\varepsilon
\end{split}
\end{equation}

Here, n($\epsilon$,V) denotes the total electronic density of states (DOS), f($\epsilon$,T) is the Fermi-Dirac distribution, $\epsilon_F$ is the Fermi energy, and k$_B$ is the Boltzmann constant.

The vibrational energy of the lattice is calculated using the quasi-harmonic Debye-Grüneisen approximation, as described in~\cite{Moruzzi1988PhysRevB}:
\begin{equation}
    F_{vib}(V,T) = E_{vib}(V,T) - TS_{vib}(V,T)
\end{equation}
The ion vibrational energy E$_{vib}$(V,T) is given by:
\begin{equation}
E_{\mathrm{vib}}(T,V) = \frac{9}{8} N_A k_B \theta_D + 3 N_A k_B T D\left(\frac{\theta_D}{T}\right)
\end{equation}
To compute the vibrational entropy S$_{vib}$(T,V), the following expression is used:
\begin{equation}
\begin{aligned}
S_{\mathrm{vib}}(T,V) 
&= 3 N_A k_B \left[ \frac{4}{3} D\left( \frac{\theta_D}{T} \right) \right. \\
&\quad \left. - \ln\left( 1 - \exp\left( - \frac{\theta_D}{T} \right) \right) \right]
\end{aligned}
\end{equation}
where D($\theta_D$/T) is the Debye function. The Debye temperature, $\theta_D$(V), is calculated using:
\begin{equation}
\theta_D(V) = \theta_{D0} \left( \frac{V_0}{V} \right)^\gamma
\end{equation}
The Grüneisen parameter $\gamma$ is obtained from the following relation:
\begin{equation}
\gamma = -1 - \frac{V}{2} \frac{\frac{\partial^2 P}{\partial V^2}}{\frac{\partial P}{\partial V}}
\end{equation}

The Debye temperature $\theta_{D0}$ is calculated at the equilibrium volume V$_0$ and T = 0 K, as listed in Table~\ref{table:wave_vel}. 

The magnetic entropy $S_{\mathrm{mag}}(V)$ can be calculated within the mean-field approximation as
\[
S_{\mathrm{mag}}(V) = k_{\mathrm{B}} \sum_{i=1}^{n} c_i \ln\left(\left|\mu_i(V)\right| + 1\right),
\label{eq:Smag}
\]
where $\mu_i$ and $c_i$ denote the local magnetic moment and the concentration of atom $i$, respectively. However, since molybdenum carbide (Mo$_2$C) is a non-magnetic material, the contribution of magnetic entropy to the free energy is considered negligible.

The theoretical analysis of anisotropic thermal expansion in crystalline materials is inherently complex because the crystal volume depends on multiple lattice parameters. In the case of orthorhombic Mo$_2$C, the volume is given by
\[
V(a,b,c) = a \cdot b \cdot c.
\]
Consequently, the evaluation of thermodynamic quantities such as pressure $P(a,b,c)$ or the bulk modulus $B(a,b,c)$ requires partial derivatives of the total energy $E(a,b,c)$ with respect to each lattice parameter, rather than simple derivatives with respect to volume, as in 
\[
P(V) = -\frac{dE}{dV} \quad \text{or} \quad B(V) = -V \frac{dP}{dV}.
\]

A further complication arises from the fact that the standard expressions for the thermodynamic free energy $F(V,T)$, including the electronic and vibrational contributions described in Eqs. (21–29), are formulated as functions of a single variable—the volume—while temperature T acts as a parameter. Strictly, applying these expressions to anisotropic systems would require reformulating the free energy as $F(a,b,c,T)$, which constitutes a mathematically nontrivial problem.

To overcome this difficulty, a simplified Search of Thermal Expansion Path (STEP) method ~\cite{Vasilyev2021PhysicaB} ~\cite{Vasilyev2023MTComm} ~\cite{Vasilyev2023PSSB} ~\cite{Vasilyev2024PCCP} ~\cite{VasilyevGorev2025PhysicaB} was employed. In the present work, the method was adapted for orthorhombic crystal lattices. The STEP approach reduces the multidimensional problem to an effective one-dimensional formulation by evaluating the total energy $E(V)$ along selected fixed directions in lattice-parameter space, thereby treating the free energy as a function of volume only. This strategy enables the direct use of conventional equations of state (e.g., Murnaghan or Vinet), the Debye model, and standard expressions for electronic and vibrational entropies, all of which depend explicitly on volume.

Within this framework, the free energies given by Eqs. (21–29) are calculated along different candidate pathways, and the direction corresponding to the minimum free energy is identified as the preferred thermal expansion path of the compound. Total energies were obtained from first-principles quantum-mechanical calculations, while thermodynamic properties were modeled using the Debye–Grüneisen approach within the quasi-harmonic approximation (QHA).

\section{Thermal expansion calculation scheme}
\begin{figure*}[t]
\centering
  \includegraphics[width=10cm]{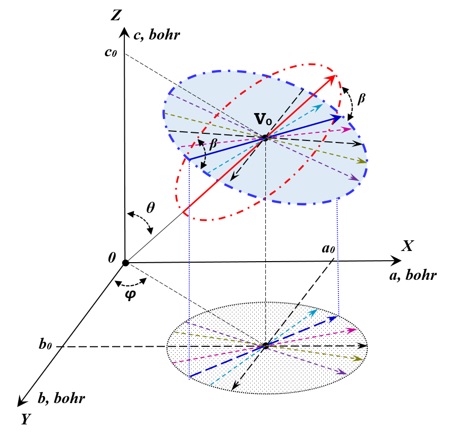}\\
  \caption 
  {Scheme for determining thermal expansion paths of orthorhombic $\alpha-$Mo$_2$C. The lattice parameters are plotted along the XYZ axes. The point V$_0$ = (a$_0$, b$_0$, c$_0$) corresponds to the optimized equilibrium volume at T = 0 K obtained from first-principles calculations. The red arrow indicates the isotropic thermal expansion/contraction direction, lying in the red plane and forming an angle $\theta$ with the OZ axis. The angles $\theta$ and $\phi$ correspond to latitude and longitude, respectively. The blue arrow represents the predicted thermal expansion direction. Multiple candidate expansion paths within the same blue plane are shown by colored arrows, and their projections onto the XOY plane are displayed below.}
\label{fig:figure3}  
\end{figure*}

The optimized lattice parameters ($a_0$, $b_0$, $c_0$) and equilibrium volume V$_0$ of the orthorhombic $\alpha-$Mo$_2$C crystal structure were obtained from first-principles calculations and are summarized in Table~\ref{table:f_Lattice_const}. In a Cartesian coordinate system (XYZ), where the lattice parameters $a$, $b$, and $c$ are aligned along the OX, OY, and OZ axes, respectively, the equilibrium state is represented by a point V$_0$ with coordinates ($a_0$, $b_0$, $c_0$), as illustrated in Figure~\ref{fig:figure3}

A vector drawn through V$_0$, indicated by a solid red line, represents the direction of isotropic thermal expansion and contraction of $\alpha-$Mo$_2$C, along which the ratios $c/a$ and $c/b$ remain constant. This vector forms an angle $\theta$ with the OZ axis, corresponding to the latitude, and the projection of this vector onto the XOY plane forms an angle $\phi$ with the OY axis, corresponding to longitude. The isotropic expansion vector lies within a plane highlighted by a red dash–dotted circle in Figure~\ref{fig:figure3} (hereafter referred to as the “red plane”), which is centered at V$_0$. This red plane forms an angle $\theta$ with the OZ axis.

An infinite number of planes can be constructed through the point V$_0$. From these, a set of planes is selected such that each plane forms an angle $\beta$ with the red plane. One such plane is shown in Figure ~\ref{fig:figure3} as a blue dash–dotted shaded circle (hereafter referred to as the “blue plane”). Each blue plane is uniquely defined by its angle $\beta$, which may take positive or negative values; the case of positive $\beta$ is illustrated in Figure~\ref{fig:figure3}. The point V$_0$ belongs simultaneously to all such planes.

Within each blue plane, multiple candidate thermal expansion paths passing through V$_0$ can be defined. Several such paths are illustrated in Figure~\ref{fig:figure3} by colored dash–dotted vectors. Their projections onto the XOY plane are shown in the lower part of the figure.

\begin{figure*}[t]
\centering
  \includegraphics[width=12cm]{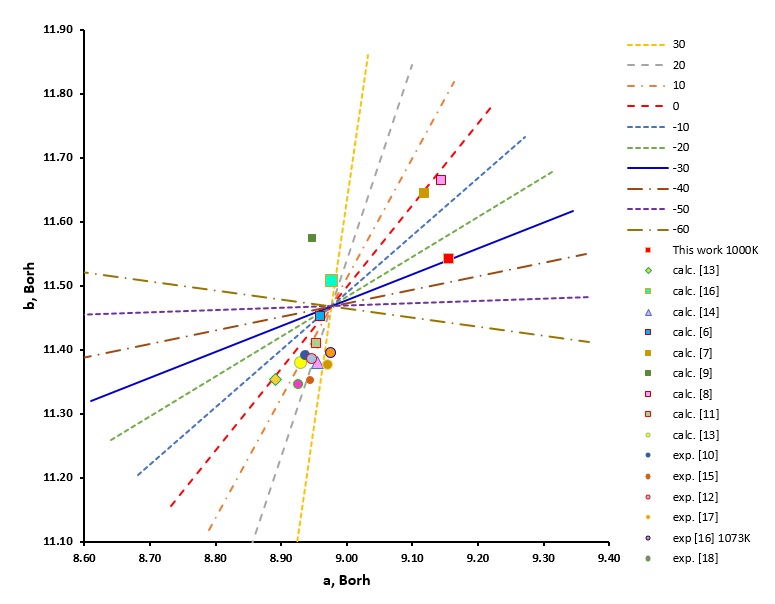}\\
  \caption{Projection of the $\beta = 60^\circ$ plane onto the XOY plane. The intersection point of all paths (30, 20, 10, 0, -10, -20, ..., -60) corresponds to the projection of $V_0$ with optimized lattice parameters ($a_0$, $b_0$, $c_0$) of orthorhombic $\alpha$-Mo$_2$C. The red dashed line (0) denotes the pseudo-isotropic thermal expansion path, where only the ${a/b}$ ratio is constant. The blue solid line (-30) represents the projection of the predicted thermal expansion path. Projections of experimental and calculated lattice parameters reported in previous studies are included for comparison.}
  \label{fig:figure4}
\end{figure*}

\begin{figure}[t]
\centering
  \includegraphics[width=8.5cm]{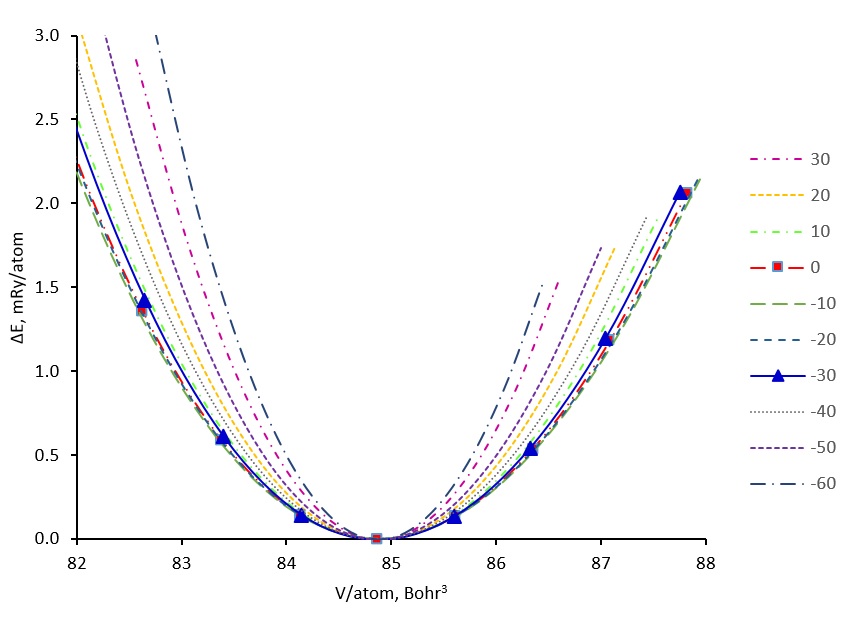} 
  \caption{Total energies $E_{\mathrm{tot}}(V)$ of orthorhombic $\alpha$-Mo$_2$C calculated from first principles as functions of volume along the candidate thermal expansion paths (30°, 20°, 10°, 0°, -10°, -20°, …, -60°) within the $\beta$-plane.}
  \label{fig:figure5}
\end{figure}

The key idea of this approach is that, when moving along such fixed directions, the thermodynamic functions $F(V,T)$ depend only on a single variable, namely the volume. Consequently, the classical expressions given in Eqs. (21–29) can be employed in their original form without additional modifications, thereby reducing the dimensionality of the problem to a one-dimensional case.

By calculating the free energies (Eqs. 21–29) along all selected directions within the chosen blue planes (characterized by different values of $\beta$) and comparing them, a unique direction corresponding to the minimum free energy can be identified. This direction represents the preferred thermal expansion and contraction path. Increasing the number of sampled directions and blue planes improves the accuracy of the predicted path and allows for the possibility that the true thermal expansion trajectory is curvilinear.

As an illustrative example, Figure~\ref{fig:figure4} shows the projection of one such blue plane with $\beta = 60^\circ$ onto the XOY plane. The set of straight lines labeled 30, 20, 10, 0, -10, -20, …, -60 corresponds to the projections of candidate thermal expansion paths lying within the $\beta = 60^\circ$ plane. All lines intersect at a common point, which is the projection of V$_0$ onto the XOY plane. The red dashed line labeled “0” represents a pseudo-isotropic thermal expansion path for which only the ratio $a/b$ remains constant. All other paths are defined relative to this reference direction, with their labels indicating the angle with respect to the pseudo-isotropic path.

The solid blue line labeled “-30” in Figure~\ref{fig:figure4} forms an angle of -30$^\circ$ with respect to the pseudo-isotropic reference direction. The sign convention for the angles is arbitrary. The present calculations indicate that the “-30” direction corresponds to the projection of the energetically favorable thermal expansion and contraction path of the $\alpha-$Mo$_2$C crystal lattice onto the XOY plane. For comparison, projections of experimentally measured and previously calculated lattice parameters of $\alpha-$Mo$_2$C are also shown in Figure~\ref{fig:figure4}.

The total energies E$_{tot}$(V) of orthorhombic $\alpha-$Mo$_2$C were calculated from first principles along the candidate thermal expansion paths 30$^\circ$, 20$^\circ$, 10$^\circ$, 0$^\circ$, -10$^\circ$, -20$^\circ$, …, -60$^\circ$ belonging to different $\beta$ planes. The total energies obtained for the set of paths lying in the “blue” plane with $\beta = 60^\circ$ are shown in Figure~\ref{fig:figure5}. Among these paths, the -10$^\circ$ direction exhibits the lowest total energy. In contrast, the total energy calculated along the predicted thermal expansion path (-30$^\circ$, shown by the blue solid line) lies within the energy spread. This result indicates that, based solely on total energy considerations, it is not possible to reliably identify the thermodynamically preferred thermal expansion path in advance.

\section{Electronic and Vibrational energy of Mo$_2$C}
\begin{figure}[t]
\centering
  \includegraphics[width=8.5cm]{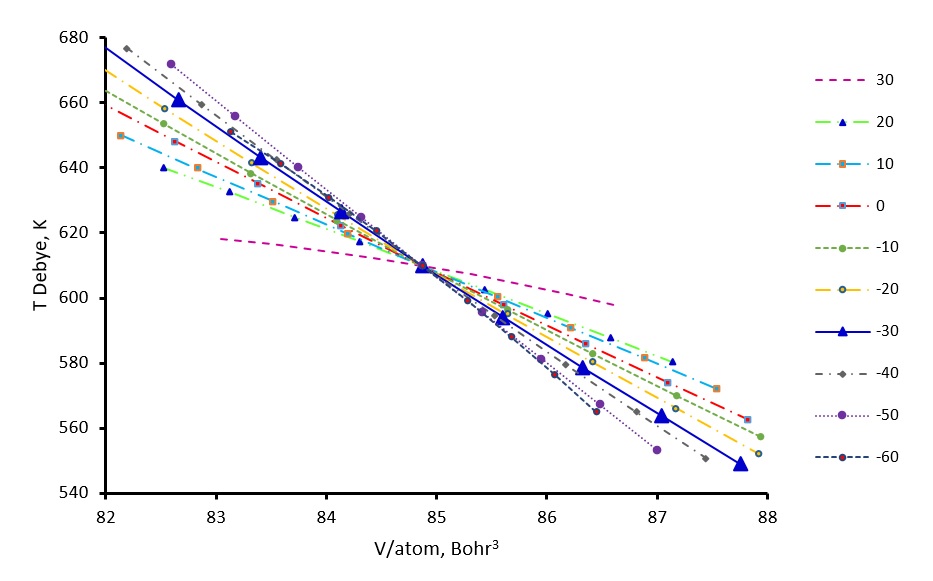}\\
  \caption{Debye temperatures $\theta$$_D$(V) of orthorhombic $\alpha-$Mo$_2$C calculated from first principles as functions of volume along the candidate thermal expansion paths (30°, 20°, 10°, 0°, -10°, -20°, …, -60°) within the $\beta$ = 60° plane.}
  \label{fig:figure6}
\end{figure}

To evaluate the vibrational contribution to the free energy, the quasi-harmonic approximation (QHA) was employed. Within this framework, the Debye temperature $\theta$$_D$(V) was calculated along the same set of paths using Eq. (28). The Debye temperatures obtained for the paths belonging to the $\beta$ = 60° plane are presented in Figure~\ref{fig:figure6}. The -50° and -60° paths exhibit the largest gradients of $\theta$$_D$(V) with respect to volume. According to the Debye–Grüneisen model~\cite{Lu2007ActaMater}, such behavior suggests a potentially larger vibrational contribution to the free energy along these directions. In contrast, the Debye temperature calculated along the 0° path (shown by the red dashed line) displays a considerably smaller slope, indicating a weaker vibrational contribution and, consequently, a lower likelihood of being energetically favorable.

The electronic contribution to the free energy was evaluated using the total electronic density of states (DOS). The DOS was calculated for all considered paths across different $\beta$ planes and subsequently used to compute the electronic free energy contribution according to Eqs. (22–24).

\section{Free energy components}
Finally, the Helmholtz free energies F(V$_0$,T) of $\alpha-$Mo$_2$C were calculated along all candidate paths using Eq. (21). The total energy E$_{tot}$(V), electronic contribution F$_{el}$(V$_0$,T), and vibrational contribution F$_{vib}$(V$_0$,T) were evaluated using Eqs. (22–29). All thermodynamic quantities were calculated at the equilibrium volume corresponding to each temperature.

\begin{figure*}[t]
\centering
  \includegraphics[width=13cm]{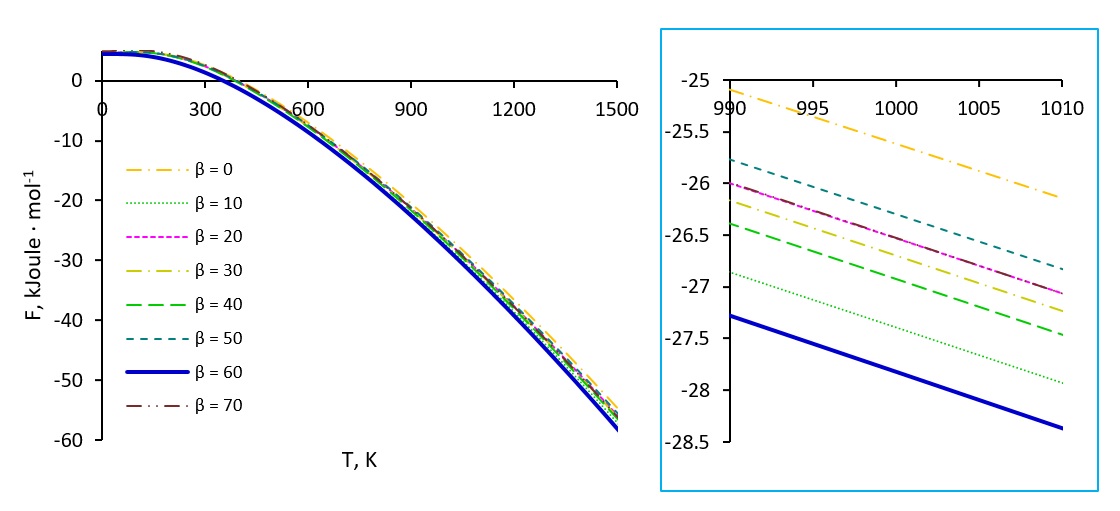}\\
  \caption{Minimum free energies of $\alpha-$Mo$_2$C calculated for different $\beta$ planes (0° to 70°). The blue solid line corresponds to the plane $\beta$ = 60°, representing the energetically preferred thermal expansion path.}
  \label{fig:figure7}
\end{figure*}
\begin{figure*}[t]
\centering
  \includegraphics[width=13cm]{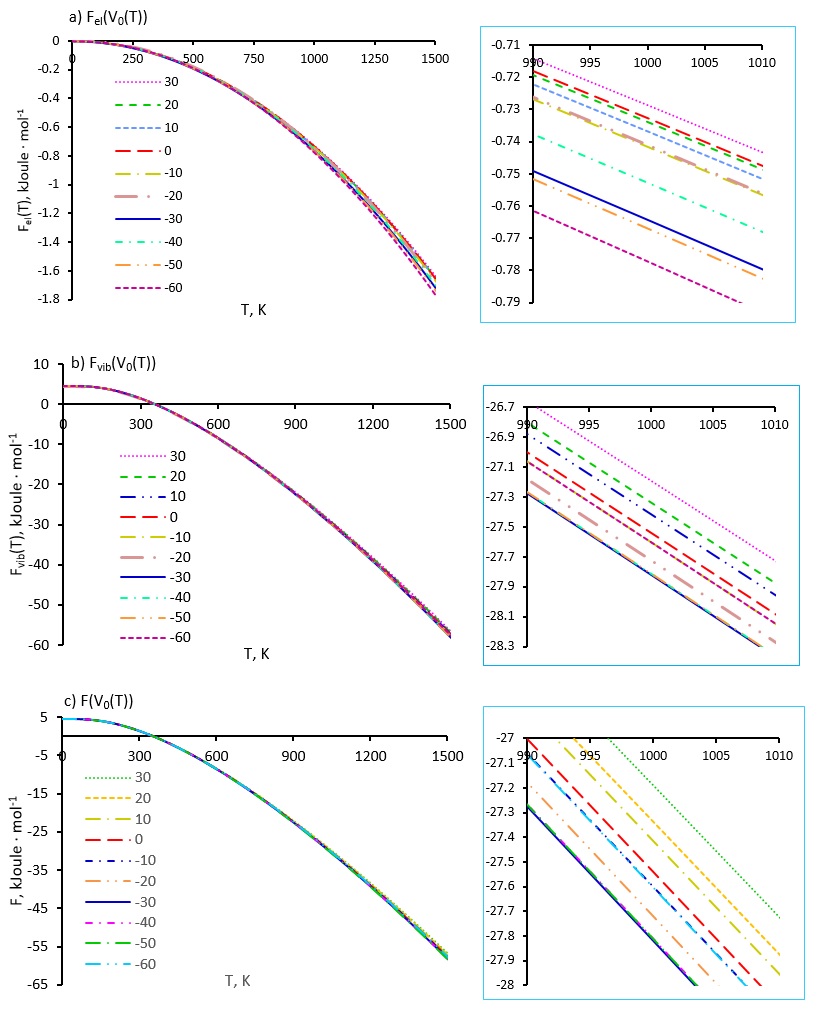}\\
  \caption{Free energies of orthorhombic $\alpha-$Mo$_2$C and their contributions along the candidate thermal expansion paths (30°, 20°, 10°, 0°, -10°, -20°, …, -60°) within the $\beta$ = 60° plane. (a) Electronic contributions, F$_{el}$(V$_0$(T)); (b) Vibrational contributions, F$_{vib}$(V$_0$(T)); (c) Total free energies, F(V$_0$(T)). Insets on the right show magnified views for clarity.}
  \label{fig:figure8}
\end{figure*}

In this study, a series of planes defined by the angle $\beta$, ranging from -20° to 70°, was investigated, where $\beta$ denotes the angle between the red and blue dash–dotted planes illustrated in Figure~\ref{fig:figure3}. For each plane, a set of ten candidate thermal expansion paths (30°, 20°, 10°, 0°, -10°, -20°, -30°, -40°, -50°, -60°) was considered, and the corresponding free energies were computed using Eqs. (22–29). Comparison of these free energies for each plane identified the energetically preferred path. The minimum free energies for eight representative $\beta$ planes are presented in Figure~\ref{fig:figure7}, indicating that the thermal expansion path of $\alpha-$Mo$_2$C lies within the plane defined by $\beta$ = 60°.

A more detailed analysis was conducted for the set of paths within the $\beta$ = 60° plane. Figure~\ref{fig:figure8}(a) shows the electronic contributions to the free energy, calculated according to Eqs. (22–24). These data indicate that the largest electronic contributions arise from paths -50° and -60°, whereas the -30° path does not correspond to the maximum electronic contribution.

The vibrational contributions to the free energy, computed from Eqs. (25–29), are presented in Figure~\ref{fig:figure8}(b). The analysis demonstrates that the -30° path provides the largest vibrational contribution, followed by paths -40° and -50°, while the -60° path contributes significantly less. This highlights that, although vibrational effects dominate the free energy, electronic contributions remain non-negligible. Neglecting the electronic component would incorrectly suggest that the -50° path is the most favorable energetically. 

The configurational entropy of the Mo$_2$C compound contributes equally to the free energy in all directions. Therefore, in a comparative analysis, it does not affect the result.

The total free energies for all paths in the $\beta$ = 60° plane are shown in Figure~\ref{fig:figure8}(c). The results indicate that the minimum free energy corresponds to the -30° path, followed by -50° and -40°. These calculations predict that the orthorhombic $\alpha-$Mo$_2$C phase will preferentially expand along the -30° direction, which lies in a plane forming an angle of $\beta$ = 60° relative to the isotropic reference direction. Given that the isotropic vector depicted in Figure~\ref{fig:figure3} has a latitude $\theta$ = 55.81°, the plane containing the preferred thermal expansion path is oriented at an angle of 115.81° with respect to the OZ axis.
\begin{figure}[t]
\centering
  \includegraphics[width=8cm]{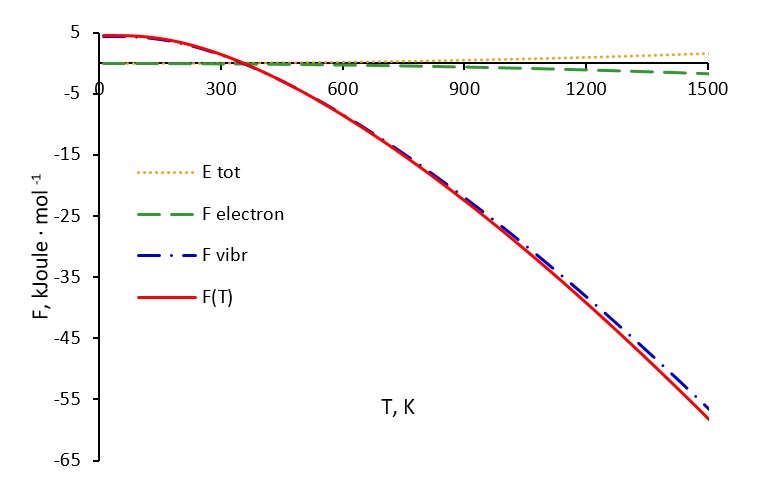}\\
  \caption{Free energy curve, F(V$_0$(T)), for $\alpha-$Mo$_2$C, showing the corresponding energy terms: total, electronic, and vibrational energies, calculated along the thermal expansion pathway with $\beta$ = 60° plane.}
  \label{fig:figure9}
\end{figure}
\begin{figure}[t]
\centering
  \includegraphics[width=8cm]{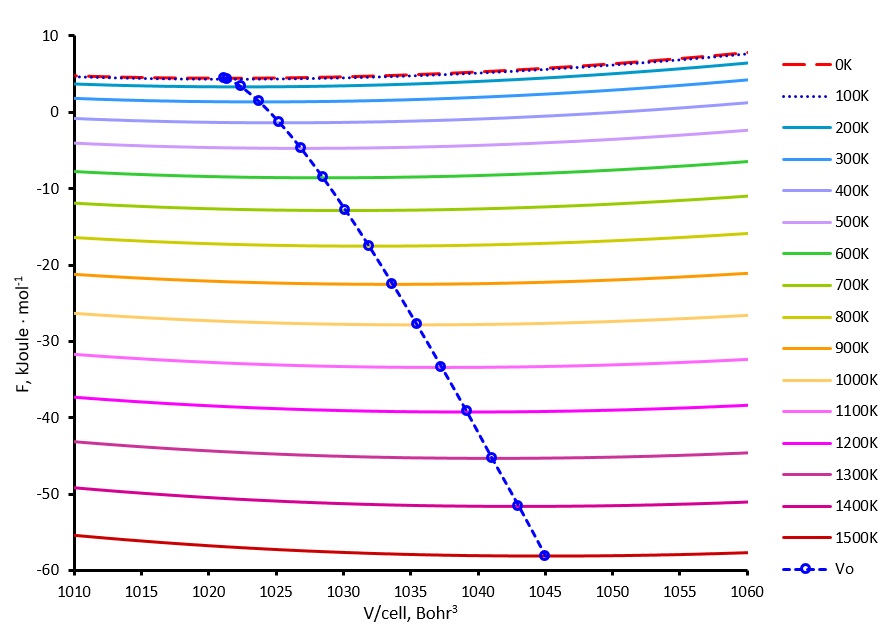}\\
  \caption{Free energy curves as a function of volume, calculated at various temperatures for $\alpha-$Mo$_2$C along the predicted thermal expansion pathway. The red dashed line represents the total energy, E$_{tot}$(V). The blue dotted line with open circles indicates the equilibrium volume, V$_0$, corresponding to the free energy at each temperature.}
  \label{fig:figure10}
\end{figure}

The energy components of free energy F(V$_0$(T)): total, electronic and vibrational energies calculated along the -30 pathway belonging to the plane $\beta$ = 60° are shown in Figure~\ref{fig:figure9}. Analysis shows that approximately ninety-seven percent of the free energy contribution comes from vibrational energy. While the energy contribution from electrons is relatively small.

The free energies curves F(V$_0$(T)) of Mo$_2$C calculated along the -30 pathway belonging to the plane $\beta$ = 60° at finite temperatures at equilibrium volumes are shown in Figure~\ref{fig:figure10}. The increase in equilibrium volume V$_0$, shown by the blue dashed line, represents the thermal expansion of the solid.

\begin{table*}[t]
\centering
 \caption{Elastic moduli (GPa) and Poisson's ratio ($\nu$) of Mo$_2$C calculated for the ground state, compared with data obtained in other studies.}
 \label{table:modulus} 
\begin{tabular}{||c c c c c c c c c c c||} 
 \hline
  & Ref. & B$_V$ & B$_R$ & $B$ & G$_V$ & G$_R$ & $G$ & $E$ & $\nu$ & $B/G$ \\ [0.5ex] 
 \hline\hline
 Calc. & Thic work & 297.27 & 297.06 & 297.17 & 168.09 & 161.26 & 164.67 & 417.00 & 0.27 & 1.80 \\ [1ex] 
 \hline
 Calc. & ~\cite{Persson2015MaterialsProject} & 294 & 294 & 294 & 152 & 148 & 150 & - & 0.28 & - \\ [1ex] 
 \hline
 Calc. & ~\cite{Karaca2019JAlloysCompd} & - & - & 305.9 & - & - & 147.9 & 382.3 & 0.29 & 2.07 \\ [1ex] 
 \hline 
 Calc. & ~\cite{Naher2022ResultsPhys} & 314.39 & 314.24 & 314.31 & 163.95 & 159.54 & 161.54 & 161.74 & 0.28 & 1.94 \\ [1ex] 
 \hline  
 Calc. & ~\cite{Politi2013PCCP} & - & - & 315.5 & - & - & - & - & - & - \\ [1ex] 
 \hline
 Exp. & ~\cite{Haines2001JPCM} & - & - & 307.5 & - & - & - & - & - & - \\ [1ex] 
 \hline
 Calc. & ~\cite{Medford2012JournalCatalysis} & - & - & 282.3 & - & - & - & - & - & - \\ [1ex] 
 \hline
\end{tabular}
\end{table*}

\section{Discussion}

\begin{figure*}[t]
\centering
  \includegraphics[width=13cm]{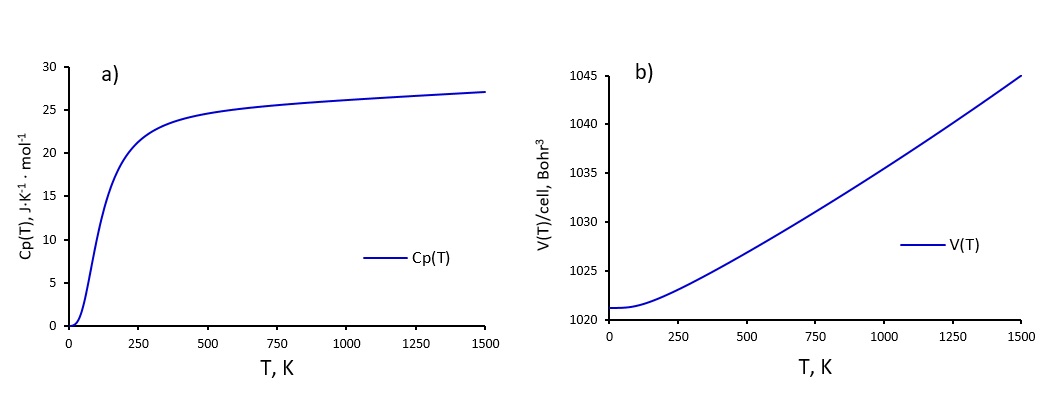}\\
  \caption{(a) Heat capacity, C$_P$(T), and (b) volumetric temperature dependence, V(T), for the orthorhombic phase of $\alpha-$Mo$_2$C, calculated along the predicted thermal expansion pathway.}
  \label{fig:figure11}
\end{figure*}

In this study, the DFT + QHA approach, combined with a modified Search of Thermal Expansion Path (STEP) method tailored to orthorhombic structures was employed to predict the thermal expansion behavior of $\alpha-$Mo$_2$C. The working diagram for the modified STEP method is presented in Figure~\ref{fig:figure3}, and the optimized lattice parameters of Mo$_2$C, computed at the ground state using DFT, are listed in Table~\ref{table:f_Lattice_const}. Additionally, Figure~\ref{fig:figure4} illustrates the projection of the equilibrium volume (V$_0$) onto the XOY coordinate plane, with experimental data for Mo$_2$C lattice parameters included for comparison. 

The calculated lattice parameters of Mo$_2$C, as presented in Table 1, exhibit a high degree of concordance with previous experimental and theoretical findings, with discrepancies not exceeding 1.5\%. However, a direct comparison with earlier theoretical studies is complicated by variations in computational methodologies, including differences in functional selections and other associated parameters. This issue also extends to the comparison with experimental data, as the present simulations were conducted at absolute zero temperature (T = 0 K), neglecting factors such as vacancies, compositional inhomogeneity, and temperature-dependent effects, all of which may contribute to the observed discrepancies. Mo$_2$C is conventionally produced from a stoichiometric blend of molybdenum powder and carbon black, by annealing under argon flow or vacuum at a temperature between 1000 and 1500 C~\cite{Shovensin1966PMMC}. Furthermore, according to the Mo-C phase diagram presented by Epicier et al.~\cite{Epicier1988ActaMet}, Velikanova et al.~\cite{Velikanova1988PMMC}, Otani et al.~\cite{Otani1995JCG} and Seigle et al.~\cite{Seigle1979MetallTransA}, the stoichiometric $\alpha-$Mo$_2$C compound resides either at the boundary or at the two-phase regions (bcc Mo + $\alpha$ or $\alpha$ + C), over the entire temperature range. In other words, these experimental phase diagrams predict that stoichiometric $\alpha-$Mo$_2$C is not a stable compound at low temperatures. These boundary compositions pose significant challenges for experimental studies, as minor compositional deviations can result in erroneous conclusions. Additionally, the thermal annealing time of the samples is crucial, as it must be sufficiently long to permit diffusion processes and allow the system to reach equilibrium. Nevertheless, the negative enthalpy of formation value (Table~\ref{table:f_Lattice_const}) and the fulfillment of the stability criterion (8) further substantiate the thermodynamic stability of the Mo$_2$C compound, confirming that the $\alpha-$Mo$_2$C phase remains stable at the ground state.

The most energetically favorable thermal expansion path for $\alpha-$Mo$_2$C, identified as the -30 pathway associated with the $\beta$ = 60° plane, is shown in Figure~\ref{fig:figure4} as a projection onto the XOY coordinate plane. The calculations indicate that the lattice parameter $a$ increases more rapidly than $b$ with rising temperature, while the $c$ parameter exhibits negative thermal expansion. This anisotropic behavior is of particular importance for understanding the material’s performance, especially in applications such as electrocatalysis. It is also highly relevant for evaluating materials used in solar cells, where thermal compatibility between perovskite layers and charge-transport materials is a crucial design consideration.

The results further demonstrate that including electronic entropy in the calculations is essential for accurately predicting the thermal expansion behavior. Omitting the electronic entropy contribution causes a shift in the predicted thermal expansion path of $\alpha-$Mo$_2$C from the -30 pathway to an alternative path (-50), as shown in Figure~\ref{fig:figure4}. Such a misjudgment could lead to incorrect predictions regarding material thermal compatibility, which is critical for applications like electrocatalysis or components in solar cells.

The isobaric heat capacity $Cp(T)$ and volumetric temperature dependence of Mo$_2$C calculated along the predicted thermal expansion direction is given in Figure~\ref{fig:figure11} (a) and (b).

The computational insights provided by this study offer valuable guidance for engineering decisions aimed at optimizing the performance and longevity of materials with orthorhombic lattice structures. These insights are particularly relevant for applications in catalysis, superconductivity, solar cell materials, advanced electronic devices, and other cutting-edge technologies.

\section{Conclusion}
The Search of Thermal Expansion Path (STEP) method has been adapted to accommodate the orthorhombic crystal structure. First-principles calculations, coupled with the quasi-harmonic Debye-Grüneisen approximation, were employed to determine the contributions of electronic and vibrational free energies to the thermal behavior of $\alpha-$Mo$_2$C. The predicted thermal expansion path is in close agreement with experimental data, confirming the anisotropic nature of the thermal expansion in this material. Specifically, the lattice parameter $a$ exhibits a more rapid increase with temperature compared to parameter $b$, while the $c$ parameter demonstrates a negative coefficient of thermal expansion (CTE). Key thermodynamic properties, including the elastic constants, bulk modulus, sound velocities, and Debye temperature, were calculated for the ground state. Furthermore, the heat capacity and volumetric temperature dependence were evaluated along the thermal expansion path of Mo$_2$C. The STEP method proves to be an effective tool for studying the thermodynamic properties of materials with orthorhombic structures. The results indicate that vibrational energy plays a dominant role in determining the stability of Mo$_2$C, while electronic entropy is another significant factor that must be accurately accounted for to reliably predict the thermal expansion anisotropy of orthorhombic $\alpha-$Mo$_2$C.

\section{Acknowledgement}
The author acknowledges the support of the Basic Research Program and HPC facilites at HSE University.

\section{Conflicts of interest or competing interests}
The author has no relevant financial or non-financial interests to disclose.

\section{Data and code availability}
There are no data that support the findings of this study (Not Applicable).

\section{Supplementary information}
There is no supplementary material used for this manuscript (Not Applicable).

\section{Ethical approval}
Not Applicable.

\bibliographystyle{elsarticle-num}
\bibliography{ref}

\end{document}